\documentclass[pra,twocolumn,amsmath,amssymb,floatfi,showpacs]{revtex4}
\usepackage{graphicx}

\begin{document}
\title{\textbf{Nonradiative interaction and entanglement between distant atoms}}
\author{Ephraim Shahmoon}
\author{Gershon Kurizki}
\affiliation{Department of Chemical Physics, Weizmann
Institute of Science, Rehovot, 76100, Israel}
\date{\today}

\begin{abstract}
We show that nonradiative interactions between atomic dipoles placed in a waveguide can give rise to deterministic entanglement at ranges much larger than their resonant wavelength. The range increases as the dipole-resonance approaches the waveguide's cutoff frequency, caused by the giant density of photon modes near cutoff, a regime where the standard (perturbative) Markov approximation fails. We provide analytical theories for both the Markovian and non-Markovian regimes, supported by numerical simulations, and discuss possible experimental realizations.
\end{abstract}

 \maketitle

\emph{Introduction.---}
Dipoles can interact via photon exchange, resulting in excitation transfer and mutual entanglement \cite{SCU}. When the interaction is mediated by radiation, i.e. real photons, it constitutes a dissipative and hence quantum-mechanically incoherent process, whereby the generation of entanglement is generally probabilistic \cite{DLCZ,POL}, although certain entangled states are deterministically obtainable by engineering/control of the bath \cite{4}. In this study, we are concerned with the nonradiative interaction that stems from the collective coupling of atomic dipoles to a common "bath" of photonic modes \cite{5}. Such nonradiative (dispersive) interactions are possible via their near or evanescent fields \cite{RON}. Quantum mechanically they are described as exchange of \emph{virtual}, i.e. non-resonant, \emph{photons} between the atoms, known as resonant dipole-dipole interaction (RDDI) \cite{MQED,LEH,MEY}. In free space RDDI is dominant over radiation only at distances shorter than the resonant wavelength. Here we predict modified RDDI along with suppressed radiation in confined geometries, giving rise to \emph{coherent} interaction at distances much longer than the resonant wavelength. This constitutes a novel route towards \emph{high-fidelity long-range deterministic entanglement}.
The principle that allows to appropriately modify the radiative and dispersive interactions is that they are mediated by the geometry-dependent field modes, populated by either real or virtual photons, respectively. Hence, the distance-dependence of the interactions is determined by the geometry. For example, when mediated by surface-plasmon-polariton modes in one dimension, both interactions appear to have long-range character, yet they are hindered by dissipation mechanisms \cite{SPA,FLE1,FLE2}. E.g., in \cite{SPA}, the radiative interaction sets the bound of the concurrence (entanglement) at $C=0.5$. This bound is circumvented by  a promising approach to a coherent phase-gate based on the difference between super- and sub-radiant decay rates \cite{FLE1}. Still, ohmic losses and radiation to free-space modes may practically limit the phase-gate operation to distances smaller than a wavelength. Radiation, however, can be suppressed in geometries that create cutoffs or bandgaps in the photonic mode spectrum. In such geometries RDDI can be drastically modified \cite{KUR,SEK,LAW}.

In our approach, photonic cutoffs or bandgaps are used not only to suppress radiation but also to enhance RDDI so as to make it the dominant effect.
Our main result, obtained by essentially exact (nonperturbative) calculations, is the possibility of extremely long-distance RDDI almost without radiation, and correspondingly high concurrence (nearly-perfect entanglement). This effect is predicted in waveguides for pairs of atoms whose dipolar transition frequency is just below the cutoff or bandedge of the waveguide. We thereby reveal the key principle that enables coherent long-range interaction, potentially much stronger than possible decoherence effects, namely, the very large density of photon states near the cutoff.
Thus, the enhancement of density of states due to the cutoff is reminiscent of that obtained using a cavity. However, unlike a cavity, the waveguide geometry is open along the propagation axis and does not restrict the separation of the atoms.
In the Markov approximation, the RDDI diminishes with the interatomic distance $z$ as $e^{-z/\xi}$, where $\xi$ increases as the atomic frequency approaches the cutoff (bandedge), allowing for entanglement at long distances. Yet, the standard Markov approximation fails close to cutoff, which requires a nonperturbative analysis, supported by numerical calculations.

\emph{The model.---} We consider a pair of atoms, modeled by identical two-level-systems (TLS) with energy levels $|g\rangle$ and $|e\rangle$ and transition frequency $\omega_a$. These are coupled to the vacuum field of a non-leaky waveguide, i.e. we neglect the TLS coupling to modes outside the waveguide -- a relevant assumption in the situation considered below.
The TLS-field dipole couplings are $g_{k\alpha}=\sqrt{\frac{\omega_k}{2\epsilon_0\hbar}}\mathbf{d}\cdot\mathbf{u}_k(\mathbf{r}_{\alpha})$,
$\mathbf{r}_{\alpha}$ being the location of atom $\alpha=1,2$, $\mathbf{d}$ the dipole matrix element of the $|g\rangle\leftrightarrow|e\rangle$ transition (taken to be real), and $\omega_k$ and $\mathbf{u}_k(\mathbf{r})$ the $k$'th mode frequency and spatial function.
The corresponding Hamiltonian in the dipole approximation \cite{CCT, MQED} reads, in the interaction picture,
\begin{equation}
H_{AF}=\hbar\sum_{\alpha=1}^2\sum_k\left[i g_{k\alpha} \hat{a}_k e^{-i\omega_k t} +h.c.\right]\left[\hat{\sigma}_{\alpha}^{-}e^{-i\omega_a t}+h.c.\right],
\label{H_AF}
\end{equation}
$\hat{a}_k$, $\hat{\sigma}^{-}_{\alpha}$ being the mode and the TLS lowering operators, respectively. In what follows, we analyze the atomic dynamics under the perturbative Markov approximation and without it.

\emph{Markovian theory.---}
Adopting an open-system approach for the problem \cite{CAR}, we identify the two atoms as the system and the continuum of EM vacuum modes as a bath, and consider the effects of the bath on the system  (Fig. 1(a)). These are dissipative and dispersive effects that are related by the Kramers-Kronig relation and determined by the bath's two-point (autocorrelation) spectrum $G_{\alpha\alpha'}(\omega)$, defined via
\begin{equation}
\sum_k g_{k\alpha}g_{k\alpha'}^{\ast}\longrightarrow \int d \omega G_{\alpha \alpha'}(\omega).
\label{G}
\end{equation}

\begin{figure}
\begin{center}
\includegraphics[scale=0.45]{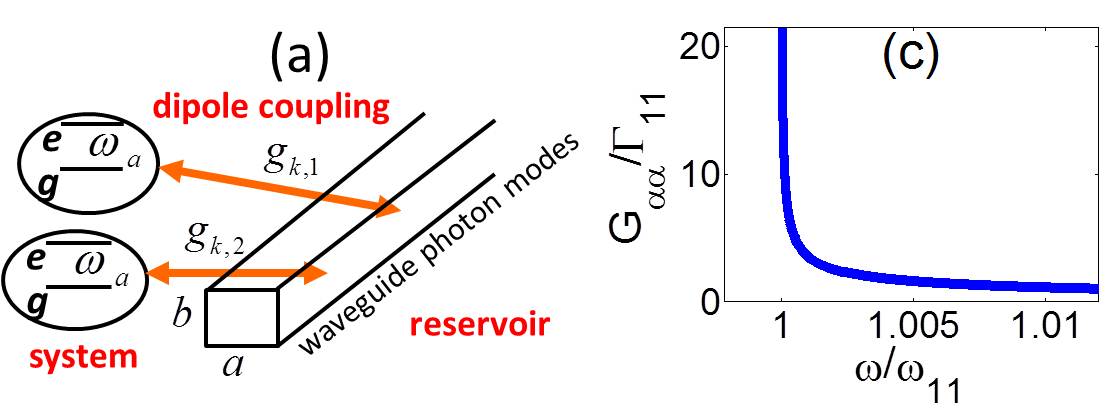}
\includegraphics[scale=0.2]{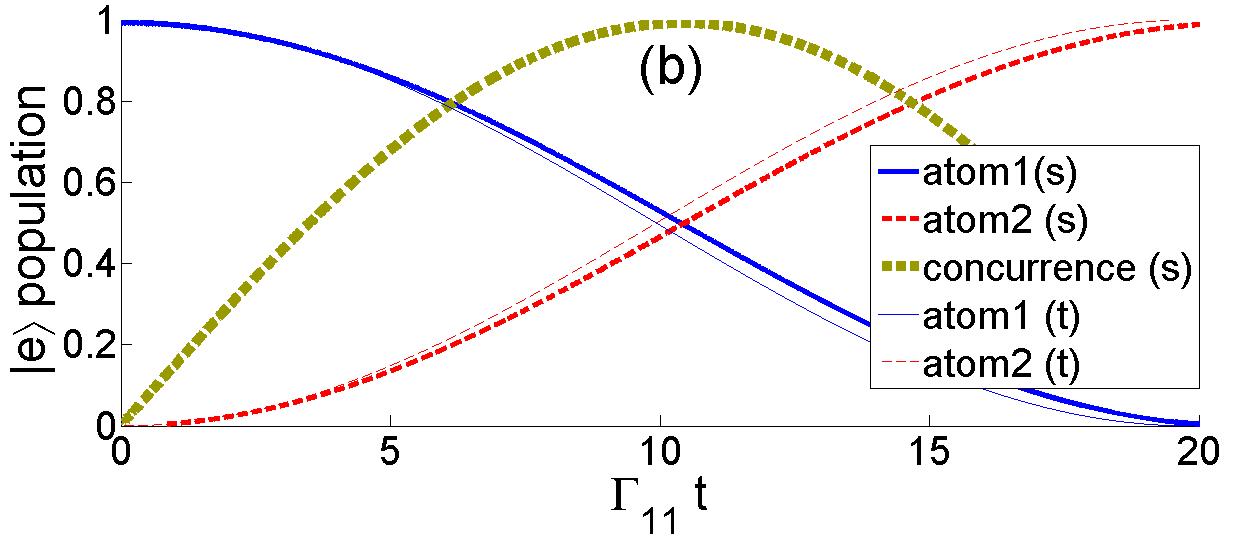}
\caption{\small{(color online). (a) Schematic picture of our model. (b) Dynamics of excited state population for the atoms affected by the $TM_{11}$ mode well below cutoff, with [Eqs. (6)-(9)] $\omega_{11}=500\Gamma_{11}$, $\omega_a-\omega_{11}=-100\Gamma_{11}$ and $z_{12}=0.5\lambda_a$: Markovian theory (t) and simulation (s) results.  (c) Divergence near cutoff $\omega_{11}$ of the single-atom spectrum $G_{\alpha \alpha}$ for $TM_{11}$ mode, Eq. (\ref{G_TM}).
 }} \label{fig1}
\end{center}
\end{figure}

From Fermi's Golden rule we obtain the rate of dissipation by radiation  $\gamma_{\alpha\alpha'}=2\pi G_{\alpha\alpha'}(\omega_a)$, which for $\alpha=\alpha'$ represents the single-atom spontaneous emission rate to the guided modes and for $\alpha\neq\alpha'$ describes the two-atom, distance-dependent, cooperative emission \cite{15}. The dispersive effect is obtained by second-order perturbation theory for the energy correction (cooperative Lamb shift \cite{16}) of the two-atom states, associated with the bath-induced dipole-dipole Hamiltonian term,
\begin{equation}
H_{DD}=-\hbar\frac{1}{2}\sum_{\alpha\alpha'}\Delta_{\alpha\alpha'}\left(\hat{\sigma}_{\alpha}^{+}\hat{\sigma}_{\alpha'}^{-}+
\hat{\sigma}_{\alpha}^{-}\hat{\sigma}_{\alpha'}^{+}\right),
\label{H_DD}
\end{equation}
where $\Delta_{\alpha\alpha'}=\Delta_{\alpha\alpha',-}+\Delta_{\alpha\alpha',+}$ and
\begin{equation}
\Delta_{\alpha\alpha',\mp}=\mathrm{P}\int_0^{\infty} d\omega\frac{ G_{\alpha\alpha'}(\omega)}{\omega\mp\omega_a},
\label{D}
\end{equation}
P denoting the principal value.
The dissipative, incoherent effect of $\gamma_{\alpha\alpha'}$ gives rise to probabilistic interaction between the atoms. Hence, in order to achieve non-radiative, deterministic interaction we need a vanishing $\gamma_{\alpha\alpha'}$, leaving intact the coherent dynamics governed by $H_{DD}$ in Eq. (\ref{H_DD}). Then, if initially only atom 1 is excited, we get a periodic exchange of the excitation between the atoms, at a rate $\Delta_{12}$, in the two-atom state
\begin{equation}
|\psi_{12}(t)\rangle=\cos(\Delta_{12} t)|e_1,g_2\rangle + i\sin(\Delta_{12} t)|g_1,e_2\rangle,
\label{psi12}
\end{equation}
that superposes singly-excited product states of atoms 1 and 2. A maximally entangled state is then achieved at odd multiples of the time $t=\pi/(4\Delta_{12})$.

In order to illustrate how the radiative effects $\gamma_{\alpha\alpha'}$ can be suppressed we first consider the case of atoms placed inside a rectangular hollow metallic waveguide (MWG), with longitudinal axis $z$ and transverse dimensions $a$ and $b$ (see Fig. 1a). Nonideal MWG and optical fiber realizations will be addressed below. The atom interacts only with the MWG field modes $TE_{mn}$ (transverse electric) and $TM_{mn}$ (transverse magnetic) labeled by non-negative integers $m,n$ \cite{KONG} (see Appendix). Each $TE/TM_{mn}$ transverse mode has its own cutoff frequency $\omega_{mn}$ and dispersion relation $\omega^{mn}_{k_z}$, $k_z$ being the longitudinal wavenumber,
\begin{eqnarray}
\omega_{mn}&=&c\sqrt{(m\pi/a)^2+(n\pi/b)^2}
\nonumber \\
\omega^{mn}_{k_z}&=&\sqrt{(c k_z)^2+\omega_{mn}^2},
\label{DR}
\end{eqnarray}
where $\omega_k=\omega^{mn}_{k_z}$ is the frequency of the $k=TE/TM_{mn,k_z}$ mode, and $c$ is the speed of light.
The contribution of a specific transverse mode  $\lambda_{mn}$ ($\lambda=TE,TM$) to the bath spectrum in Eq. (\ref{G}) is obtained from the dispersion relation $k_z(\omega)$ (Eq. (\ref{DR})) upon identifying $\omega^{mn}_{k_z}=\omega$,
\begin{eqnarray}
G^{\lambda}_{mn,\alpha\alpha'}(\omega)&=&\frac{\partial k_z}{\partial \omega} g^{\lambda}_{mn, \alpha}(\omega) g^{\lambda \ast}_{mn, \alpha'}(\omega)\Theta(\omega-\omega_{mn})
\\ \label{Gmn}
\frac{\partial k_z}{\partial \omega}&=&\frac{1}{c}\frac{\omega}{\omega_{mn}}\frac{1}{\sqrt{(\omega/\omega_{mn})^2-1}},
\label{DOS}
\end{eqnarray}
 $\Theta(x)$ being the Heaviside step function. At this stage two key features of the waveguide structure must be noted: \emph{(1)} below the cutoff $\omega_{mn}$ no $\lambda_{mn}$ guided photon modes exist, and \emph{(2)} the density of states $\frac{\partial k_z}{\partial \omega}$ diverges near the cutoff. In what follows, we use feature \emph{(1)} to suppress radiation and feature \emph{(2)} to obtain long-distance and strong RDDI.

In order to facilitate the analysis it is sufficient to consider the case where the atoms are polarizable only in the $z$ direction, $\mathbf{d}=d_z \mathbf{e}_z$ (for other polarizations see the Appendix or Ref. \cite{LAW}). Since $TE$ modes have a vanishing $z$ component of the electric field, only $TM$ modes contribute to the bath spectrum,
\begin{equation}
G_{\alpha\alpha'}(\omega)=\sum_{mn}\frac{\Gamma_{mn}}{2\pi}\frac{\cos\left[k_z(z_{\alpha}-z_{\alpha'})\right]}{\sqrt{(\omega/\omega_{mn})^2-1}}\Theta(\omega-\omega_{mn}).
\label{G_TM}
\end{equation}
Here $\Gamma_{mn}\equiv\frac{4 \omega_{mn}\tilde{d}^{(z)}_{mn,\alpha}\tilde{d}^{(z)}_{mn,\alpha'}}{\pi\epsilon_0\hbar c a b}$ is introduced, where  $\tilde{d}^{(z)}_{mn,\alpha}=d_z\sin\left(\frac{m\pi}{a}x_{\alpha}\right)\sin\left(\frac{n\pi}{b}y_{\alpha}\right)$ and  $x_{\alpha},y_{\alpha}$ is the transverse position of atom $\alpha$. Also note that $k_z$ is a function of $\omega$ by virtue of Eq. (\ref{DR}).

Now, consider the case where the atomic resonance is below the lowest cutoff frequency, $\omega_a<\omega_{11}$ for $TM$ modes. Then, the atomic dipoles are not resonant with any of the field modes and radiation is suppressed, $\gamma_{\alpha \alpha'}=2 \pi G_{\alpha\alpha'}(\omega_a)=0$, from Eq. (\ref{G_TM}). We are thus left only with the nonradiative RDDI Eq. (\ref{D}),
\begin{equation}
\Delta_{12}=\sum_{mn}\frac{\Gamma_{mn}}{2}\frac{1}{\sqrt{1-(\omega_a/\omega_{mn})^2}}e^{-\frac{z_{12}}{\xi_{mn}}},
\label{D12}
\end{equation}
where $z_{12}\equiv |z_1-z_2|$ and the effective interaction range is
\begin{equation}
\xi_{mn}=\frac{c}{\omega_{mn}}\frac{1}{\sqrt{1-(\omega_a/\omega_{mn})^2}}.
\label{xi}
\end{equation}
These Markovian-theory results \cite{LAW} predict that radiative dissipation is absent, while the RDDI decays exponentially with interatomic distance, typical of interaction mediated by evanescent waves. Yet, remarkably, Eqs. (\ref{D12}) and (\ref{xi}) imply that as the atomic resonance $\omega_a$ approaches the lowest cutoff $\omega_{11}$ from below, the RDDI diverges owing to the contribution of the $TM_{11}$ mode, and so does its range, determined by $\xi_{11}$. This potentially enables deterministic generation of entanglement at very large distances.

 In order to test the above results we performed direct numerical simulations of the Schr\"{o}dinger equation for the Hamiltonian (\ref{H_AF}), taking only the dominant $m=1,n=1$ mode into account (see Appendix). Fig. 1(b) portraits typical dynamics of the atoms' populations, along with their entanglement, quantified by the concurrence \cite{WOO}, for $z_{12}=0.5\lambda_a$ with $\lambda_a$ the atomic transition wavelength. As expected from Eq. (\ref{psi12}), the maximally entangled state is generated at half the oscillation period of the population exchange. It is also apparent that when $\omega_a$ is not too close to the cutoff $\omega_{11}$, the simulation results agree with those of the Markovian analysis, Eqs. (\ref{psi12}) and (\ref{D12}), within numerical accuracy.

\emph{Validity of the Markovian theory.---}
The Markov approximation used above breaks down as $\omega_a$ approaches the cutoff. The general conditions for the validity of the Markov approximation reduce in  our case to (see Appendix)
\begin{equation}
\Delta_{12}(\omega_a)\Delta''_{12}(\omega_a)\ll 1,
\label{MAR}
\end{equation}
where $\Delta_{12}(\omega_a)$ is given by Eq. (\ref{D12}) and $\Delta''_{12}(\omega_a)$ is its second derivative w.r.t $\omega_a$. In the limit $\omega_a\rightarrow\omega_{mn}$, $\Delta_{12}(\omega_a)$ and $\Delta''_{12}(\omega_a)$ become singular and condition (\ref{MAR}) is not satisfied, as seen from Eq. (\ref{D12}) and Fig. 1(c). Thus, a non-Markovian theory is required in order to fully analyze the possibility of long-distance RDDI and entanglement.
Non-Markovian analysis has been performed before for a single atom coupled to a continuum with a cutoff \cite{KOF,KNG}, yielding the possibility of incomplete decay: decay of the excited state population to a steady-state value different from zero, as a result of the formation of atom-photon bound-states.
Nevertheless, the Markovian analysis is very useful for RDDI in cases where nearly-complete entanglement (e.g. $C>0.95$) is to be achieved, as seen below.

\emph{Non-Markovian theory.---}
 In order to account for the situation where $\omega_a$ approaches the cutoff, we develop a nonperturbative and non-Markovian theory for RDDI, in the spirit of \cite{KOF}. From Hamiltonian (\ref{H_AF}), assuming that only atom 1 is initially excited, the state of the combined (atoms+modes) system can be written within the rotating-wave-approximation \cite{CCT} as
\begin{equation}
|\psi(t)\rangle=a_1(t)|e_1,g_2,0\rangle+a_2(t)|g_1,e_2,0\rangle+\sum_k b_k(t)|g_1,g_2,1_k\rangle.
\nonumber
\end{equation}
Inserting this state into the Schr\"{o}dinger equation, we obtain dynamical equations for $a_1(t)$, $a_2(t)$ and $b_k(t)$. As before, we consider only the MWG transverse mode $m=1,n=1$, this time for $\omega_a$ close to the cutoff $\omega_{11}$, such that the denominator of the spectrum (\ref{G_TM}) is well approximated by $\sqrt{(\omega/\omega_{11})^2-1}\approx\sqrt{2}\sqrt{\omega/\omega_{11}-1}$.
Using the Laplace transform in order to solve the dynamical equations, we then obtain the dynamics of the first atom (more details can be found in the Appendix),
\begin{equation}
a_1(t)=\sqrt{i}e^{-i\omega_{11}t}\sum_{j=1}^5c_j\left[\frac{1}{\sqrt{\pi t}}+\sqrt{i}u_je^{iu_j^2t}\mathrm{erfc}(-\sqrt{i}u_j\sqrt{t})\right].
\label{a1}
\end{equation}
Here $u_j$ are the roots of $d(u)=u^5+2W_au^3-\frac{1}{\sqrt{2}}\Gamma_{11}\sqrt{\omega_{11}}u^2+W_a^2u-\frac{1}{\sqrt{2}}\Gamma_{11}\sqrt{\omega_{11}}W_a-\frac{1}{8}\Gamma_{11}^2\omega_{11}\frac{1}{u}F(u)$, where $W_a=\omega_a-\omega_{11}$, $c_j=n(u_j)/d'(u_j)$ with $n(u)=-i(u^3+W_au-\frac{1}{2\sqrt{2}}\Gamma_{11}\sqrt{\omega_{11}})$, and
$F(u)=\left(e^{-2z_{12}(\sqrt{\omega_{11}}/c) u}-1\right)$, where $F(u)$ is expanded in Taylor series up to 5th order in $u$ (Appendix).
The conditions of validity for this theory are thus given by the approximation of the spectrum and the expansion of $F(u)$, yielding $\frac{\omega_a-\omega_{11}}{4\omega_{11}}\ll 1$ and $z_{12}\ll(\frac{45}{4})^{1/6}\frac{1}{2\pi} \frac{\omega_a}{\omega_{11}}\sqrt{\frac{\omega_{11}}{2|\omega_{11}-\omega_a|}}$, respectively. However, in practice, another limitation on the precision of the theory comes from the numerical calculation of the roots of $d(u)$.

Fig. 2(a) presents the dynamics of the atomic populations and interatomic entanglement in the non-Markovian regime. Very good agreement between the above theory and numerical simulations is observed. The main feature of the dynamics are Rabi-like oscillations similar to those of the Markovian case. Nevertheless, their amplitude is decreased as a result of excitation loss to the field modes by incomplete decay, setting the upper bound on the achievable entanglement. Hence, as $\omega_a$ approaches the cutoff, while the inter-atomic distance $z$ is kept fixed, we get a tradeoff between increased RDDI strength and decreased maximum entanglement. This is shown in Fig. 2(b), where $\omega_a$ is varied from very far from cutoff, where Markovian theory predictions $\Delta_{12,M}$ and $C_{max}=1$ apply, to very close to the cutoff, where $\Delta_{12}$ increases on the expense of $C_{max}$.

\begin{figure}
\begin{center}
\includegraphics[scale=0.35]{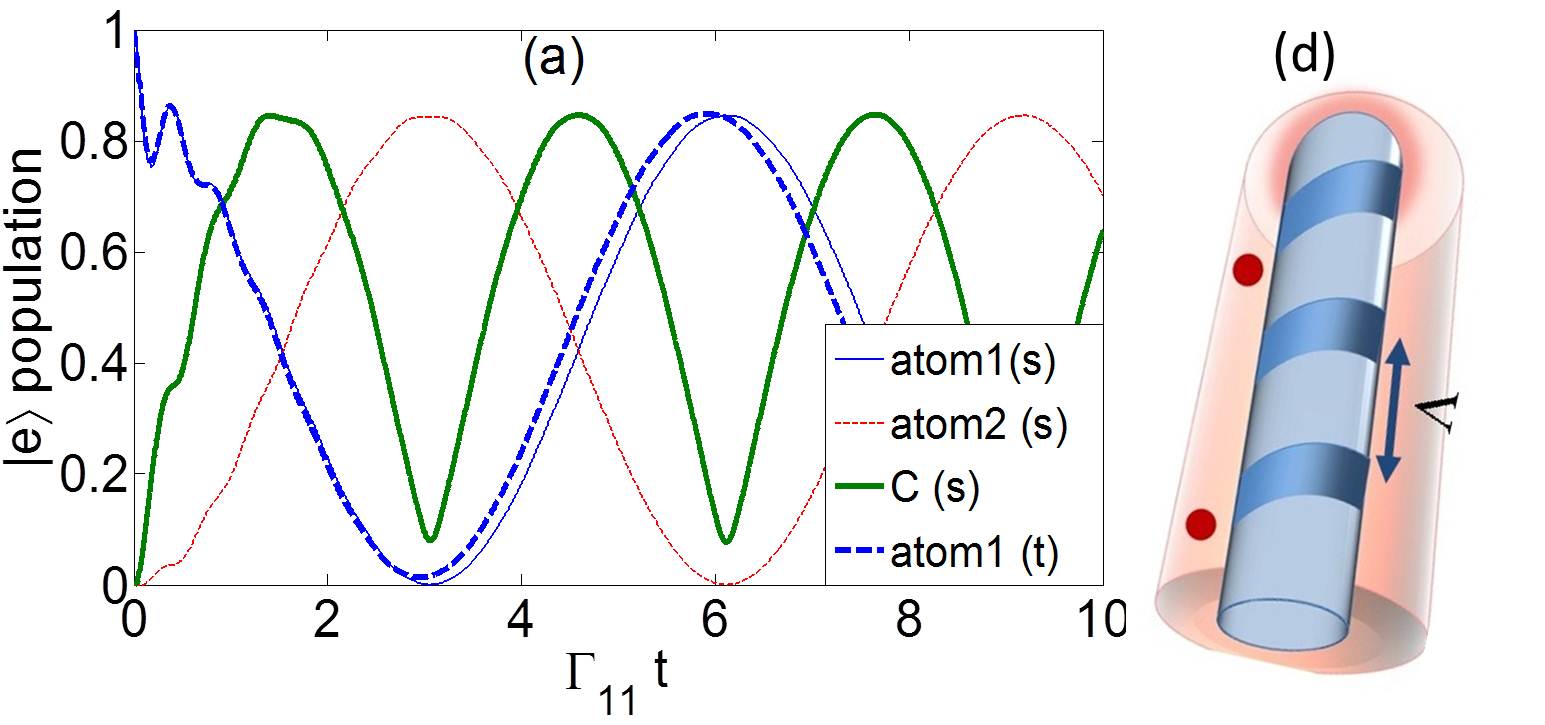}
\includegraphics[scale=0.24]{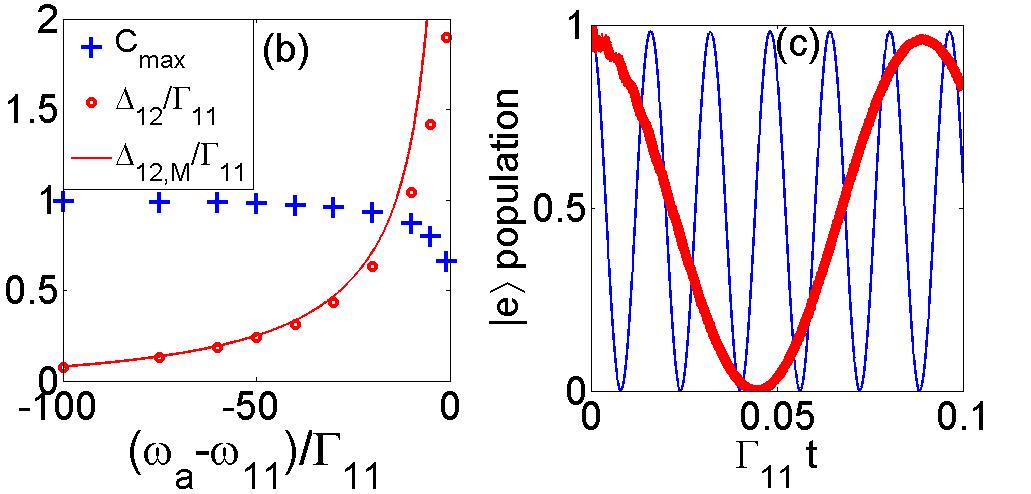}
\caption{\small{
(color online). Atomic dynamics affected by $TM_{11}$ mode in the non-Markovian regime. (a) Atomic excitation probabilities and concurrence as a function of time. Here $\omega_{11}=500\Gamma_{11}$,  $z_{12}=\lambda_a$ and $\omega_a-\omega_{11}=-10\Gamma_{11}$. The simulations (s) well agree with the theory (t), Eq. (\ref{a1}). (b) Tradeoff between RDDI strength $\Delta_{12}$ and maximal achievable concurrence $C_{max}$ as a function of the atomic-resonance mismatch with cutoff (simulation results for $\omega_{11}=500\Gamma_{11}$,  $z_{12}=0.5\lambda_a$), compared with Markovian theory $\Delta_{12,M}$, Eq. (\ref{D12}). The estimates for $\Delta_{12}$  are extracted by fitting the dynamics of simulations for various $\omega_a$ values to Eq. (\ref{psi12}).
(c) Entanglement generation as a function of time at long-distance: the excitation of the first atom is calculated by Eq. (\ref{a1}) and the concurrence is bounded by the maximal value of the plot. MWG realization (blue thin line): $\omega_{11}=2.17\times10^{10} \Gamma_{11}$, $\omega_a-\omega_{11}=-2\times10^4 \Gamma_{11}$, $z=100\lambda_a$. Fiber-Bragg-grating realization (red thick line): $\omega_{11}=6\times10^7 \Gamma_{11}$, $\omega_a-\omega_{11}=-1500 \Gamma_{11}$, $z=20\lambda_a$. (d) Fiber-Bragg-grating scheme: the atoms are coupled to the guided modes by their transverse evanescent tails \cite{RAU} or when inserted into the fiber \cite{LUK}.
}} \label{fig2}
\end{center}
\end{figure}

\emph{Long-distance entanglement and possible realizations.---} Using the analytical theory above, we shall now illustrate the possibility of long-distance entanglement by two examples. First, consider Rydberg atoms that pass through a cold metallic waveguide (MWG), similar to the setup in \cite{HAR1,HAR2} where the MWG replaces the superconducting cavity. The states $|g\rangle$ and $|e\rangle$ are the two circular states with principal quantum numbers $51$ and $50$, with transition frequency and dipole moment $\omega_a=2\pi\times51.1$GHz and $d\sim1250 e a_0$ respectively, with $e$ the charge of an electron and $a_0$ the Bohr radius \cite{HAR2}. Near the cutoff, $\Gamma_{11}$ is similar to the free-space  $|e\rangle\rightarrow|g\rangle$ decay rate, estimated to be $\frac{\omega_a^3d^2}{3\pi\epsilon_0 \hbar c^3}\approx 14.7$Hz. The corresponding dynamics for $z=100\lambda_a$ are plotted in Fig. 2(c), where $\lambda_a\sim6$mm is the atomic wavelength, such that we obtain entanglement with concurrence $C=0.983$, at a distance $z\sim 0.6$m and for interaction time $t\approx0.2$ms [Fig. 2(c)].
Considering possible imperfections we derived the dissipation rate due to ohmic losses of the atom-induced evanescent fields in a square waveguide ($a=b$), $\gamma_{loss}\leq \frac{2R_s}{\mu_0 a}$, with $\mu_0$ the vacuum permeability and $R_s$ the surface resistance (see Appendix). Normal metals may limit the achievable entanglement distance and fidelity as in \cite{FLE1}. However, for niobium superconducting plates at temperature $T<1$K, we take, as in \cite{HAR1}, $R_s=75$n$\Omega$, yielding, for $a\approx6$mm, $\gamma_{loss}=19.89$Hz, much slower than the $0.2$ms required for entanglement. Such a temperature also ensures that the thermal photon occupancy at $\omega_a$ is negligible. In addition, as analyzed in \cite{CHEN}, surface roughness of the metal plate may slightly change the mode structure and the location of the cutoff frequency, and correspondingly the calculated RDDI rate. Nevertheless, a cutoff below which the modes become evanescent with diverging density of states persists, hence the principle of our scheme still applies. Regarding our initial assumption of isolated waveguide modes, we recall that $\omega_a$ is much smaller than the typical plasma frequency in metals ($\sim 10^{16}$Hz), so that the isolated modes of a perfect-conductor used here, are indeed adequate.

Another option is that of optical fiber modes coupled to the atoms \cite{RAU,LUK}. Although the fiber's guided modes also possess cutoffs, they lack the two important features that we have highlighted for the MWG: \emph{(1)} below cutoff the atoms are coupled to outside modes, hence spontaneous emission exists at a rate comparable to that in free space; \emph{(2)} the group velocity $\frac{\partial \omega}{\partial k_z}$ does not vanish at the fiber cutoff so that the density of states $\frac{\partial k_z}{\partial \omega}$ does not diverge. We can restore the second feature by considering a fiber-Bragg-grating \cite{FBG}: then, for a transverse fiber mode with dispersion $\omega(k_z)$, the group velocity does vanish at the bandedge of the $\omega$ spectrum corresponding to $k_z=\pi/(\Lambda \bar{n})$, with $\Lambda$ the period of the grating and $\bar{n}$ the average refractive index [Fig. 2(d)]. The dispersion near the upper boundary of the gap, $\omega_u$, can be approximately written as $\omega\approx \omega_u+B(k_z-\pi/(\Lambda\bar{n}))^2$ with constant $B$, so that $\frac{\partial k_z}{\partial \omega}\propto 1/\sqrt{\omega-\omega_u}$ diverges at $\omega_u$ in the same way assumed in our non-Markovian theory (see Appendix for more details). Then, the atom can still emit to outside modes, but just below the bandedge $\omega_u$, RDDI, which is mediated by evanescent waves in the gap, can become much stronger and more long-distance, due to the divergence. We consider optical atomic transitions, e.g. the D2 line of $^{87}Rb$ atoms with $\lambda_a\approx780$nm and natural linewidth $2\pi \times6.07$MHz. The results for $z=20\lambda_a\sim16\mu$m are plotted in Fig. 2(c), yielding concurrence $C=0.9605$ after $t\approx3.55$ns of interaction.

\emph{Conclusions.---}
To conclude, the main result of this study is the demonstration of the possibility of long-distance interaction between dipoles by a nonradiative, deterministic and coherent process (RDDI) that is crucially dependent on the waveguide geometry. The proposed scheme relies mostly on the possibility of vanishing group velocity, i.e. diverging density of states, for the guided modes, at a frequency cutoff (or bandgap) of the waveguide.
An important innovation of this work is the derivation of a non-perturbative analytic theory for RDDI near a cutoff of the photonic spectrum. The theory exhibits non-Markovian features, particulary population loss of the atoms by incomplete decay and the resulting reduction of entanglement, in agreement with numerical simulations.

Possible manifestations of the predicted effect include high-concurrence entanglement as well as energy transfer between dipoles at giant separations.
The analysis and the potential realizations discussed above suggest that the effect is significant for a wide range of atomic and waveguide parameters, constrained only by the tradeoff between interaction strength and the maximal achievable entanglement.

We acknowledge the support of DIP, ISF and the Wolfgang Pauli Institute (E.S.).

\section*{APPENDIX}

\subsection{Dipole-dipole interaction for arbitrary oriented dipoles}
In the main  text we considered the case where the dipoles are oriented in the $z$ direction. For a general orientation, we need to consider all the $TE/TM_{mn,k_z}$ modes with their normalized spatial functions \cite{KONG},
\begin{widetext}
\begin{eqnarray}
&&\mathbf{u}^{TM}_{mn,k_z}(x,y,z)=\frac{2}{\sqrt{AL}}e^{ik_zz}\left(\frac{\omega_{mn}}{\omega^{mn}_{k_z}}\sin\left(\frac{m\pi}{a}x\right)\sin\left(\frac{n\pi}{b}y\right)\mathbf{e}_z
\right. \nonumber \\ &&\left.
+\frac{i k_z c}{\omega_{mn} \omega^{mn}_{k_z}}\left[c\frac{\pi}{a}m\cos\left(\frac{m\pi}{a}x\right)\sin\left(\frac{n\pi}{b}y\right)\mathbf{e}_x
    +c\frac{\pi}{b} n \sin\left(\frac{m\pi}{a}x\right)\cos\left(\frac{n\pi}{b}y\right)\mathbf{e}_y \right]\right)
\nonumber \\
&&\mathbf{u}^{TE}_{mn,k_z}(x,y,z)=\frac{2}{\sqrt{AL}}e^{ik_zz} \left[-c\frac{\pi}{b}n\cos\left(\frac{m\pi}{a}x\right)\sin\left(\frac{n\pi}{b}y\right)\mathbf{e}_x
    +c\frac{\pi}{a} m \sin\left(\frac{m\pi}{a}x\right)\cos\left(\frac{n\pi}{b}y\right)\mathbf{e}_y \right],
\nonumber \\
\label{Auk}
\end{eqnarray}
where $A=ab$ is the transverse area of the waveguide. Inserting these mode functions into Eq. (\ref{DR}), we obtain the bath spectrum,
\begin{eqnarray}
&&G_{\alpha\alpha'}(\omega)=G^{TM}_{\alpha\alpha'}(\omega)+G^{TE}_{\alpha\alpha'}(\omega)
\nonumber \\
&&G^{TM}_{\alpha\alpha'}(\omega)=\frac{1}{\pi\epsilon_0\hbar c A}\sum_{mn}\frac{\omega_{mn}}{\sqrt{(\omega/\omega_{mn})^2-1}}\left\{\cos\left[k_z(z_{\alpha}-z_{\alpha'})\right]2\tilde{d}^{(z)}_{mn,\alpha}\tilde{d}^{(z)}_{mn,\alpha'}
\right.\nonumber \\ &&\left.
+ \cos\left[k_z(z_{\alpha}-z_{\alpha'})\right]2\tilde{d}^{TM}_{mn,\alpha}\tilde{d}^{TM}_{mn,\alpha'} \left[\left(\frac{\omega}{\omega_{mn}}\right)^2-1\right]
\right.\nonumber \\ &&\left.
+ \sin\left[k_z(z_{\alpha}-z_{\alpha'})\right]2\left[\tilde{d}^{z}_{mn,\alpha}\tilde{d}^{TM}_{mn,\alpha'}- \tilde{d}^{TM}_{mn,\alpha}\tilde{d}^{z}_{mn,\alpha'}\right]\sqrt{\left(\frac{\omega}{\omega_{mn}}\right)^2-1}\right\}\Theta(\omega-\omega_{mn})
\nonumber \\
&&G^{TE}_{\alpha\alpha'}(\omega)=\frac{1}{\pi\epsilon_0\hbar c A}\sum_{mn}\frac{\omega^2}{\sqrt{\omega^2-\omega_{mn}^2}}\cos\left[k_z(z_{\alpha}-z_{\alpha'})\right]2\tilde{d}^{TE}_{mn,\alpha}\tilde{d}^{TE}_{mn,\alpha'}\Theta(\omega-\omega_{mn}),
\nonumber \\
\label{AG_MWG}
\end{eqnarray}
where $\Theta(x)$ is the Heaviside step function. The effective dipole moments read
\begin{eqnarray}
\tilde{d}^{(z)}_{mn,\alpha}&=&d_z\sin\left(\frac{m\pi}{a}x_{\alpha}\right)\sin\left(\frac{n\pi}{b}y_{\alpha}\right)
\nonumber \\
\tilde{d}^{TM}_{mn,\alpha}&=&d_x \frac{c\frac{\pi}{a} m}{\omega_{mn}} \cos\left(\frac{m\pi}{a}x_{\alpha}\right)\sin\left(\frac{n\pi}{b}y_{\alpha}\right)
+d_y \frac{c\frac{\pi}{b} n}{\omega_{mn}} \sin\left(\frac{m\pi}{a}x_{\alpha}\right)\cos\left(\frac{n\pi}{b}y_{\alpha}\right)
\nonumber \\
\tilde{d}^{TE}_{mn,\alpha}&=&-d_x \frac{c\frac{\pi}{b} n}{\omega_{mn}} \cos\left(\frac{m\pi}{a}x_{\alpha}\right)\sin\left(\frac{n\pi}{b}y_{\alpha}\right)
+d_y \frac{c\frac{\pi}{a} m}{\omega_{mn}} \sin\left(\frac{m\pi}{a}x_{\alpha}\right)\cos\left(\frac{n\pi}{b}y_{\alpha}\right),
\nonumber \\
\label{Ad}
\end{eqnarray}
with $d_j=\mathbf{d}\cdot\mathbf{e}_j$ and $x_{\alpha},y_{\alpha}$ the transverse position of atom $\alpha$.
In order to find the RDDI $\Delta_{\alpha\alpha'}=\Delta_{\alpha\alpha',-}+\Delta_{\alpha\alpha',+}$, we recall Eq. (\ref{D}),
and find by contour integration methods,
\begin{eqnarray}
&&\Delta_{12}=\Delta_{12}^{TM}+\Delta_{12}^{TE}
\nonumber \\
&&\Delta_{12}^{TM}=\sum_{mn}\frac{2\omega_{mn}}{\epsilon_0 \hbar c A}\left[\frac{1}{\sqrt{1-\frac{\omega_a^2}{\omega_{mn}^2}}}\tilde{d}^{(z)}_{mn,1}\tilde{d}^{(z)}_{mn,2}
-\sqrt{1-\frac{\omega_a^2}{\omega_{mn}^2}}\tilde{d}^{TM}_{mn,1}\tilde{d}^{TM}_{mn,2}
\right. \nonumber \\ &&\left.
+\mathrm{sign}(z_1-z_2)\left(\tilde{d}^{(z)}_{mn,1}\tilde{d}^{TM}_{mn,2}-\tilde{d}^{TM}_{mn,1}\tilde{d}^{(z)}_{mn,2}\right)\right]e^{-\frac{|z_1-z_2|}{\xi_{mn}}}
\nonumber \\
&&\Delta_{12}^{TE}=\sum_{mn}\frac{2\omega_{mn}}{\epsilon_0 \hbar c A}\frac{\omega_a^2}{\omega_{mn}^2}\frac{1}{\sqrt{1-\frac{\omega_a^2}{\omega_{mn}^2}}}\tilde{d}^{TE}_{mn,1}\tilde{d}^{TE}_{mn,2}e^{-\frac{|z_1-z_2|}{\xi_{mn}}},
\label{AD12}
\end{eqnarray}
with $\xi_{mn}=\frac{c}{\omega_{mn}}\frac{1}{\sqrt{1-(\omega_a/\omega_{mn})^2}}$.
\end{widetext}

\subsection{Numerical simulations}
We performed direct numerical simulations of the Schr\"{o}dinger equation for the Hamiltonian from Eq. (1), taking only the dominant $TM_{11}$ mode into account. The dipole couplings $g_{k}$ relate to the 1d spectrum, from Eq. (7), by $g_{\omega,\alpha}=\sqrt{G_{\alpha\alpha}(\omega)d\omega}e^{ik_z z_{\alpha}}$, where $d\omega$ is the sampling resolution used to discretize the frequency space $\omega$. The initial atomic state is $|e_1,g_2\rangle$ where the modes are in the vacuum $|0\rangle$. By taking the rotating wave approximation \cite{CCT}, i.e. neglecting non-energy-conserving Hamiltonian terms of the form $\hat{\sigma}^{+}\hat{a}_{\omega}^{\dag},\hat{\sigma}^{-}\hat{a}_{\omega}$, we restrict ourselves to the single-excitation Hilbert space, $|e_1,g_2,0\rangle$, $|g_1,e_2,0\rangle$ and $\{|g_1,g_2,1_{\omega}\rangle,\forall \omega\}$, which is solved numerically.

\subsection{Validity of the Markov approximation}
The dissipative and dispersive coefficients, $\gamma_{\alpha\alpha'}$ and $\Delta_{\alpha\alpha'}$, can be obtained by deriving the master equation \cite{CCT,CAR} for the atoms' density matrix. Equivalently, here we will use instead the latter, second order perturbation theory for the transition amplitude. We begin with Eq. (23) on page 28 of Ref. \cite{CCT},
\begin{widetext}
\begin{equation}
U_{\alpha\alpha'}^{(2)}=
\frac{1}{2\pi i}\int_{-T/2}^{T/2}dt_1\int_{-T/2}^{T/2}dt_2\int_{-\infty}^{\infty}d\omega e^{i(\omega_a-\omega)(t_2-t_1)}W_{\alpha\alpha'}(\omega),
\end{equation}
where $U^{(2)}_{\alpha\alpha'}$ is the second order contribution to the transition amplitude from the state where only atom $\alpha$ is excited to the state where only atom $\alpha'$ is excited, $T$ is the interaction time, and
\begin{equation}
W_{\alpha\alpha'}(\omega)=
\mathrm{lim}_{\eta\rightarrow0^+}\left[\sum_k\frac{g_{k\alpha}g^{\ast}_{k\alpha'}}{\omega-\omega_k-i\eta}+\sum_k\frac{g_{k\alpha'}g^{\ast}_{k\alpha}}{\omega-2\omega_a-\omega_k-i\eta}\right].
\end{equation}
Recalling the definition of the bath spectrum in Eq. (\ref{G}), we can rewrite $W_{\alpha\alpha'}$ as
\begin{equation}
W_{\alpha\alpha'}(\omega)=\mathrm{lim}_{\eta\rightarrow0^+}\left[\int d\omega'\frac{G_{\alpha\alpha'}(\omega')}{\omega-\omega'-i\eta}+\int d\omega'\frac{G_{\alpha\alpha'}(\omega')}{\omega-2\omega_a-\omega'-i\eta}\right].
\end{equation}
Using the relation $\mathrm{lim}_{\eta\rightarrow0^+}\frac{1}{x+i\eta}=i\pi\delta(x)+\mathrm{P}\frac{1}{x}$ under integration, we obtain
\begin{equation}
U_{\alpha\alpha'}^{(2)}=\frac{1}{2\pi i}\int_{-\infty}^{\infty}d\omega \delta_T^2(\omega-\omega_a)\left[-i\frac{1}{2}\gamma_{\alpha\alpha'}(\omega)-i\frac{1}{2}\gamma_{\alpha\alpha'}(\omega-2\omega_a)-\Delta_{\alpha\alpha'}(\omega)-\Delta_{\alpha\alpha'}(\omega-2\omega_a)\right],
\label{AU}
\end{equation}
with $\delta_T(\omega)=\int_{-T/2}^{T/2}dt e^{-i\omega t}$ being a sinc function of width $1/T$ and amplitude $T$, and
\begin{equation}
\gamma_{\alpha\alpha'}(\omega)=2 \pi G_{\alpha\alpha'}(\omega)
\:\: ; \:\:
\Delta_{\alpha\alpha'}(\omega)=\mathrm{P}\int d\omega'\frac{G_{\alpha\alpha'}(\omega')}{\omega'-\omega}.
\end{equation}
In the limit $T\rightarrow\infty$, i.e. $\delta_T(\omega)\sim\delta(\omega)$, we recover the Markovian results $\gamma_{\alpha\alpha'}=\gamma_{\alpha\alpha'}(\omega_a)$ and $\Delta_{\alpha\alpha'}=\Delta_{\alpha\alpha'}(\omega_a)+\Delta_{\alpha\alpha'}(-\omega_a)$ [noting that $G_{\alpha\alpha'}(\omega<0)=0$]. Let us specify when such a limit is reasonable. Consider $T$ as the time-resolution we are interested in, i.e. $T$ is much smaller than the typical time-scale of the atomic dynamics. Nevertheless, we assume that $T$ is sufficiently large, such that in a width $1/T$ of $\delta_T^2(\omega-\omega_a)$ around $\omega_a$, $\gamma_{\alpha\alpha'}(\omega),\Delta_{\alpha\alpha'}(\omega)$ do not change appreciably. Then, we can expand $\gamma_{\alpha\alpha'}(\omega),\Delta_{\alpha\alpha'}(\omega)$ around $\omega_a$ (and also around $-\omega_a$ for $\Delta_{\alpha\alpha'}$) and get
\begin{equation}
\int_{-\infty}^{\infty}d\omega \delta_T^2(\omega-\omega_a)\Delta_{\alpha\alpha'}(\omega)\propto \Delta_{\alpha\alpha'}(\omega_a)+O\left( \frac {\Delta''_{\alpha\alpha'}(\omega_a)}{T^2}\right),
\end{equation}
\end{widetext}
where a similar result is obtained for $\gamma_{\alpha\alpha'}$. For the Markovian approximation to be valid, we demand that the lowest order relative correction for the Markovian result is small,
\begin{equation}
\frac{\Delta''_{\alpha\alpha'}(\omega_a)}{\Delta_{\alpha\alpha'}(\omega_a)}\frac{1}{T^2}\ll 1.
\label{Ac3}
\end{equation}
As a typical atomic dynamics time-scale, for the case of RDDI, we may take $\Delta_{\alpha\alpha'}$. Then, using it in (\ref{Ac3}), we obtain the condition of validity in Eq. (\ref{MAR}).

\subsection{Non-Markovian theory}
Taking the Laplace transform of the dynamical equations for $a_1(t)$, $a_2(t)$ and $b_k(t)$ with the initial conditions
$a_1(0)=1, \: a_2(0)=b_k(0)=0$, we find
\begin{equation}
\tilde{a}_1(s)=\left[s+J_{11}(s)+i\omega_a-\frac{J_{12}(s)J_{21}(s)}{s+J_{22}(s)+i\omega_a}\right]^{-1},
\label{Aas}
\end{equation}
Here $\tilde{a}_1(s)$ is the Laplace transform of $a_1(t)$ and
$J_{\alpha \alpha'}(s)=\sum_k\frac{g^{\ast}_{k,\alpha}g_{k,\alpha'}}{s+i\omega_k}$. We note that by virtue of Eq. (\ref{D}), $J_{\alpha \alpha'}(-i\omega_a)=-i\Delta_{\alpha \alpha',-}$.
As before, we consider the spectrum in Eq. (\ref{G_TM}) for $m=1,n=1$. Since $\omega_a$ is close to the cutoff $\omega_{11}$, the main contribution to RDDI comes from frequencies near $\omega_{11}$ so that we approximate the denominator of the spectrum by $\sqrt{(\omega/\omega_{11})^2-1}\approx\sqrt{2}\sqrt{\omega/\omega_{11}-1}$.
After performing the integrals in $J_{\alpha\alpha'}(s)$, using the approximated spectrum, we obtain
\begin{eqnarray}
\tilde{a}_1(s)&=&\tilde{a}_1(u)=\frac{n(u)}{d(u)}
\nonumber \\
n(u)&=&-i\left(u^3+W_a u-\frac{1}{2\sqrt{2}}\Gamma_{11}\sqrt{\omega_{11}}\right)
\nonumber \\
d(u)&=&u^5+2W_a u^3-\frac{1}{\sqrt{2}}\Gamma_{11}\sqrt{\omega_{11}}u^2
W_a^2u-
\nonumber \\
&&\frac{1}{\sqrt{2}}\Gamma_{11}\sqrt{\omega_{11}}W_a-\frac{1}{8}\Gamma_{11}^2\omega_{11}\frac{1}{u}F(u),
\nonumber \\
\label{Aau}
\end{eqnarray}
with $u=\sqrt{-i}\sqrt{s+i\omega_{11}}$, $W_a=\omega_a-\omega_{11}$ and
$
F(u)=\left(e^{-2(z_1-z_2)(\sqrt{\omega_{11}}/c) u}-1\right)
$.
In order to perform the inverse Laplace transform we first expand $F(u)$ in a Taylor series: in order to still satisfy the Laplace initial value theorem, $\alpha_1(t=0^+)=\lim_{s\rightarrow \infty}s\tilde{\alpha}_1(s)$, the expansion is taken up to 5th order. Then, expanding $\tilde{a}_1(u)$ in partial fractions \cite{KOF},
\begin{equation}
\tilde{a}_1(u)=\sum_{j=1}^5\frac{c_j}{u-u_j}
\:\: ; \:\:
c_j=c(u_j) \:\: ; \:\: c(u)=\frac{n(u)}{d'(u)},
\label{APF}
\end{equation}
where $u_j$ are the roots of $d(u)$, and using the inverse transform of $1/(\sqrt{s}+a)$ \cite{ABR}, we finally obtain
\begin{equation}
a_1(t)=\sqrt{i}e^{-i\omega_{11}t}\sum_{j=1}^5c_j\left[\frac{1}{\sqrt{\pi t}}+\sqrt{i}u_je^{iu_j^2t}\mathrm{erfc}(-\sqrt{i}u_j\sqrt{t})\right].
\label{Aa1}
\end{equation}

\subsection{Metal waveguide realization: ohmic losses}
We consider ohmic losses on the four conducting plates that make up the waveguide. The dissipated power per unit area of a plate is given by
\begin{equation}
dP_{loss}/dS=0.5|J_s|^2R_s,
\label{APs}
\end{equation}
where $S$ is the area, $R_s$ its surface resistance \cite{ORF}. In order to find the surface current $J_s$ we should first find the electric field of the dipole inside the waveguide. Assuming, as before, that the dipole is oriented to the $z$ direction, its field is a superposition of evanescent $TM_{mn}$ modes of a single $\omega_a<\omega_{mn}$ photon,
\begin{equation}
\mathbf{E}_{mn}(\mathbf{r})=i\sqrt{\frac{\hbar \omega_a}{2\epsilon_0}}\mathbf{u}^{TM}_{mn,\omega_a}(\mathbf{r}),
\label{AE}
\end{equation}
where $\mathbf{u}^{TM}_{mn,\omega_a}(\mathbf{r})$ is given by Eq. (\ref{Auk}) with $\kappa=(1/c)\sqrt{\omega_{mn}^2-\omega_a^2}$ replacing $-ik_z$ and $2\kappa$ replacing $1/L$. We then find the corresponding magnetic field using the Maxwell equations for TM modes \cite{KONG,ORF},
\begin{equation}
\mathbf{H}_{mn}(\mathbf{r})=-\frac{c^2}{\omega_{mn}^2}i\omega_a\epsilon_0\mathbf{\nabla}_{\perp}\times(\mathbf{E}_{mn}\cdot\mathbf{e}_z),
\label{H}
\end{equation}
$\mathbf{\nabla}_{\perp}=\partial_x\mathbf{e}_x+\partial_y\mathbf{e}_y$ being the curl operator in the $xy$ plane. The surface currents on the plates are found from the surface boundary conditions for the magnetic fields, $\mathbf{J}_s=\mathbf{e}_n\times\mathbf{H}$, with $\mathbf{e}_n$ the normal to the surface. Finally we integrate Eq. (\ref{APs}) over the plate area, e.g., for the plate at $y=b$, $P_{loss}=2\int_0^{\infty}dz\int_0^a dx 0.5|J_s|^2R_s$. By defining the dissipation rate per $TM_{mn}$ mode as $\gamma^{mn}_{loss}=P_{loss}/(\hbar\omega_a)$, we find for all four plates,
\begin{equation}
\gamma^{mn}_{loss}=\left[\frac{2}{\left(\frac{m}{n}\frac{b}{a}\right)^2+1}\right]\frac{R_s}{\mu_0b}+
\left[\frac{2}{\left(\frac{n}{m}\frac{a}{b}\right)^2+1}\right]\frac{R_s}{\mu_0 a}.
\label{Aloss}
\end{equation}
Then, for the case $a=b$, the total dissipation of a single photon field from the atom is bounded by $2\frac{R_s}{\mu_0 a}$.

\subsection{Fiber-Bragg-grating realization}
We briefly show how we can relate the fiber-Bragg-grating case to the theory derived for the MWG in the main text. The dispersion of a transverse fiber-mode with a Bragg-grating is \cite{FBG},
\begin{equation}
\omega(k_z)-\omega_B=\pm \frac{1}{2}\frac{\Delta n}{\bar{n}}\omega_B\sqrt{1+\left(\frac{2}{\Delta n}\right)^2\left(\frac{k_z}{k_ B}-1\right)^2},
\end{equation}
where $k_B=\omega_B/c=\pi/(\Lambda \bar{n})$ is the Bragg wavevector, $\Lambda$ the grating period, $\bar{n}$ the average refractive index and $\Delta n$ the index difference of the grating. Near the upper cutoff of the bandgap, $k_z$ is close to $k_B$ and we approximate the dispersion as
\begin{equation}
\omega(k_z)\approx\omega_u+B(k_z-k_B)^2,
\end{equation}
where $\omega_u=\omega_B(1+0.5\Delta n/\bar{n})$ is the upper bandedge and $B=\left(\frac{c}{\bar{n}}\right)^2\left(\frac{\bar{n}}{\Delta n}\right)\frac{1}{\omega_B}$. Then, the density of states is
\begin{equation}
\frac{\partial k_z}{\partial \omega}\approx\frac{\bar{n}}{c}\sqrt{\frac{\bar{n}}{4\Delta n}}\frac{1}{\sqrt{(\omega/\omega_u)-1}},
\label{ADP}
\end{equation}
where $\omega_B\approx\omega_u$ was taken. There are three terms on the right-hand-side of Eq. (\ref{ADP}): the first is a linear dispersion contribution of a mode with group velocity $c/\bar{n}$, while the second increases the usual density of states by a constant factor. The third term is the divergence due to the bandedge. The spectrum of the fiber mode will then have the form [see Eq. (7)]
\begin{equation}
G_{\alpha\alpha}(\omega)\sim\frac{\Gamma_u}{2\pi}\frac{1}{\sqrt{(\omega/\omega_u)-1}},
\end{equation}
where $\Gamma_u$ is similar to the free space spontaneous emission rate. This is the spectrum assumed in our non-Markovian theory for the MWG, with $\omega_u,\Gamma_u$ replacing $\omega_{11},\Gamma_{11}$.

\end{document}